\documentclass[showpacs,twocolumn,preprintnumbers,amsmath,amssymb,amsfonts,prb,floatfix]{revtex4}

\usepackage{graphicx, color,rotating, textcomp}
\usepackage{bm}

\begin{document}

\title{Graphene Nanogap for Gate Tunable Quantum Coherent Single Molecule Electronics}

\author{A. Bergvall}
\author{K. Berland}
\author{P. Hyldgaard}
\author{S. Kubatkin}
\author{T. L\"ofwander}
\email{tomas.lofwander@chalmers.se}

\affiliation{Department of Microtechnology and Nanoscience - MC2,\\
Chalmers University of Technology, SE-412 96 G\"oteborg, Sweden}

\date{August 23, 2011}

\begin{abstract}

  We present atomistic calculations of quantum coherent electron
  transport through fulleropyrrolidine terminated molecules bridging a
  graphene nanogap.
  We predict that three difficult problems in molecular electronics
  with single molecules may be solved by utilizing graphene contacts:
  (1) a back gate modulating the Fermi level in the graphene leads
  facilitate control of the device conductance in a transistor effect
  with high on/off current ratio;
  (2) the size mismatch between leads and molecule is avoided, in
  contrast to the traditional metal contacts;
  (3) as a consequence, distinct features in charge flow patterns
  throughout the device are directly detectable by scanning
  techniques.
  We show that moderate graphene edge disorder is unimportant for the
  transistor function.

\end{abstract}

\pacs{73.63.-b, 73.63.Rt, 85.65.+h, 72.80.Vp}

% 73.63.-b	Electronic transport in nanoscale materials and structures
% 73.63.Rt	Nanoscale contacts
% 85.65.+h	Molecular electronic devices
% 72.80.Vp	Electronic transport in graphene

\maketitle

\section{Introduction}

The isolation of graphene\cite{nov04} has lead to a wide range of
scientific discoveries and technological
opportunities.\cite{Geim2007,Geim2009} These include, for instance, a
half-integer quantum Hall effect\cite{nov2005,zhang2005} with
potential for a drastically improved quantum resistance
standard,\cite{tza2010} and a high electron mobility in this
atomically thin crystal for applications in radiofrequency
electronics\cite{lin2010,liao2010} with reduced short channel
effects\cite{sch2010} or as a transparent flexible
electrode.\cite{bae2010} In a wide perspective, graphene is a material
with potential as a versatile and controllable bridge between the
atomic and the micron scales, with unique opportunities for
nanoelectronics applications. Chemistry tools may alter graphene
properites either globally (for example chemical
gating\cite{lara2011}) or locally (atoms and molecules binded to
graphene\cite{schedin2007,can2011}). At the same time, modern
lithographic techniques compatible with semi-conducting technology can
be used to pattern graphene into devices and integrate them with
conventional electronics.\cite{pal2010} Building on these discoveries,
we show in this paper how graphene patterned to form a nanogap can be
used as electrodes for gate-tunable molecular electronics with single
molecules. The transistor effect is achieved by gating the graphene
electrodes while additional device functionalities can be built into
the molecule bridging the nanogap.

The idea of utilizing single molecules as active elements for
electronics applications are based on a number of
observations,\cite{Joa_review,Cuniberti_book,Bjo_review,cuevas_book}
including device minituarization, reproducibility, and functionality.
Besides being the ultimately small object, molecules can be
mass-replicated and their functionality can be tailored through
molecular synthesis. Many functional single molecule devices have been
demonstrated,\cite{Tao_review} but many problems remain before
practical applications can be realized. Traditionally, a metal such as
gold has been used to make contacts, although other configurations
such as semiconducting substrates combined with the scanning tunneling
microscope\cite{gui2004,rak2004} (STM), have been shown to work as
well. The huge size mismatch between metallic leads ($>10$ nm
thick) and molecules is unavoidable, which makes nanogap fabrication
challenging. Utilization of the STM for contacting molecules is not a
scalable technology and mainly suited for making devices for
research. In addition to the problem of making nanogaps, an equally
important problem for molecular electronics is how to make good
molecular transistors.\cite{kub03,oso08} The difficulty is to put a
gate sufficiently close to the molecule in a metallic nanogap to
achieve a gate effect.

Here, we study theoretically the prospects of using graphene as a
platform for single molecule electronics, where graphene
nanostructures are used as contacts and interconnects instead of metal
wires. The main purpose of this paper is to show that a transistor
effect can be achieved by utilizing a back gate that changes the
electron density and the density of states at the Fermi level of the
graphene leads. This transistor effect works well when the coupling of
the gate to the molecule is {\it weak} compared with the coupling of
the gate to the leads, which is the opposite situation to a
traditional molecular transistor, including nanotube-based
devices\cite{guo2006} although sharing the same robustness as these
through a resonance tunneling mechanism.\cite{hyld} In addition to the
transistor effect, the usage of graphene, being one-atom thick, would
circumvent the size-mismatch problem experienced with metal contacts.

The current fast improvements in graphene patterning and device
fabrication,\cite{biro2010,wang10,he2010} have opened new
opportunities for making advanced devices. These include gas
sensors,\cite{schedin2007} nanopores for DNA
sequencing,\cite{postma2010,pra2011} and single-electron transistors
operated as read out devices.\cite{can2011} In the latter experiment,
magnetic molecules were deposited on top of a graphene
constriction. By utilizing an external magnetic field, the spin states
of the molecules were manipulated, as could be read off by the
graphene single-electron transistor working in its conducting
state. We conclude that with this rapid progress in graphene device
fabrication, the devices we shall study here can be made in the near
future.

\section{Model}

\begin{figure}
  \includegraphics[width=\columnwidth]{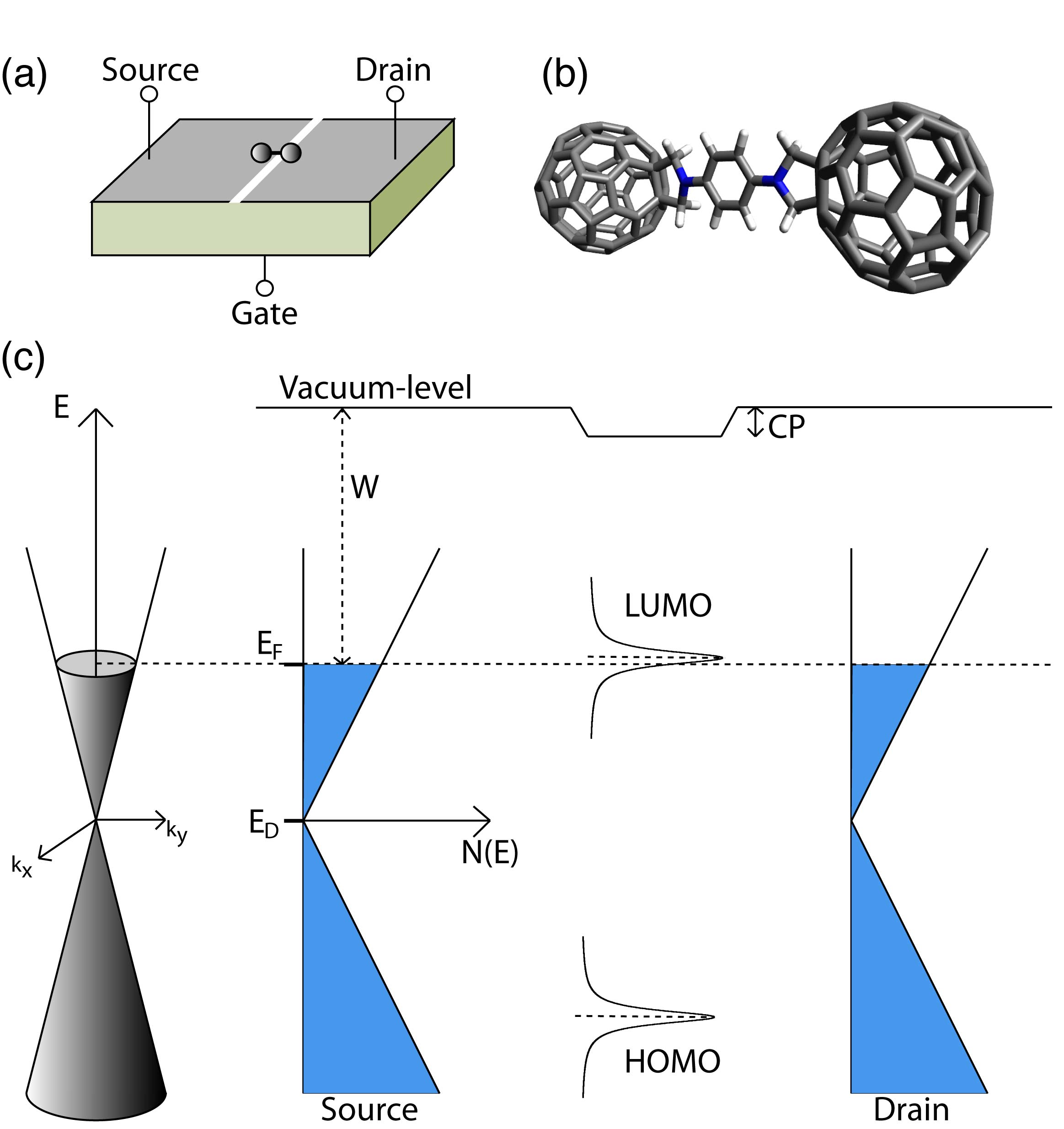}
  \caption{(a) Geometry of the transistor. (b) {\it Dumbbell} molecule
    consisting of a 1,4-phenylenediamine bridge with C60 anchoring
    groups. (c) Energy level diagram at zero source-drain voltage.
    The work function $W$ of the leads can be changed by a few hundred
    meV via a back gate so that the Dirac point $E_D$ in the graphene
    density of states $N(E)$ is above or below (depicted) the Fermi
    level $E_F$. The estimated charge transfer to the molecule,
    reflected in the contact potential $CP$, is such that the Fermi
    level intersects the LUMO.}
\label{Fig-ELD}
\end{figure}

The geometry of the molecular transistor we are considering is shown
in Fig.~\ref{Fig-ELD}(a).  The molecule, resembling a dumbbell,
consists of a central wire of 1,4-phenylenediamine with C60 end-groups
(i.e. fulleropyrrolidine terminated benzene), as depicted in
Fig.~\ref{Fig-ELD}(b). The wide graphene leads extends far from the
molecule and are electrically connected to source and drain. A back
gate can be used to change the position of the Dirac point in the
graphene bandstructure with respect to the molecular energy levels. We
describe this gate effect in more detail below.

The goal of this work is not to predict the functionality of a certain
molecule in great detail, or to reproduce or explain a certain
experiment. Rather, the goal is to show the most salient features of a
single-molecule transistor with graphene leads operating in the
quantum coherent regime. But we base our studies on the specific
molecule in Fig.~\ref{Fig-ELD}. This molecule has recently attracted a
lot of attention because it has promising properties for making
reproducible contacts via the large C60 anchoring
groups.\cite{BDC_zant,BDC_JOC} Thus, both leads and anchoring groups
are made of carbon.

We shall in this work use a minimal model based on a tight-binding
description. The Hamiltonian is
\begin{equation}
H = \sum_i E_i c_i^{\dagger}c_i + \sum_{i \neq j} t_{ij} c_i^{\dagger}c_j,
\label{hamiltonian}
\end{equation}
where $E_i$ are onsite energies, $t_{ij}$ are hopping amplitudes
between sites $j$ and $i$, and the operators $c_i^{\dagger}$ and $c_i$
create and destroy electrons on sites $i$. We concentrate on the
quantum coherent transport regime and leave effects of Coulomb
interaction to future studies. This corresponds to assuming a small
charging energy $U$ on the molecule compared with the molecular level
broadening $\Gamma$ due to the coupling to the leads. The graphene
nanostructured leads, as well as the molecule, are readily built up by
restricting the sites $i$ and $j$.  We study both armchair and zigzag
graphene nanoribbon leads with nearest neightbor hopping amplitude
$t$, with and without edge disorder. We include large and wide
sections of the graphene leads in the calculations and connect them to
ideal ribbons connected to reservoirs through the technique of
self-energies. This is a Landauer approach based on non-equlibrium
Green's functions.\cite{cuevas_book,datta_book} We will for simplicity
focus on the low-bias regime and present results for the transmission
function of the device, as well as spectral charge current flow
patterns inside the device. We note that it is important to have wide
ribbons, with edge disorder on a length scale smaller than the ribbon
width, otherwise weak links may form at necks of the imperfect
graphene ribbon. Two such necks define a quantum dot with
single-electron transistor properties,\cite{ihn2010} which leads to
unwanted Coulomb blockade effects in the leads.

\begin{figure}
  \includegraphics[width=\columnwidth]{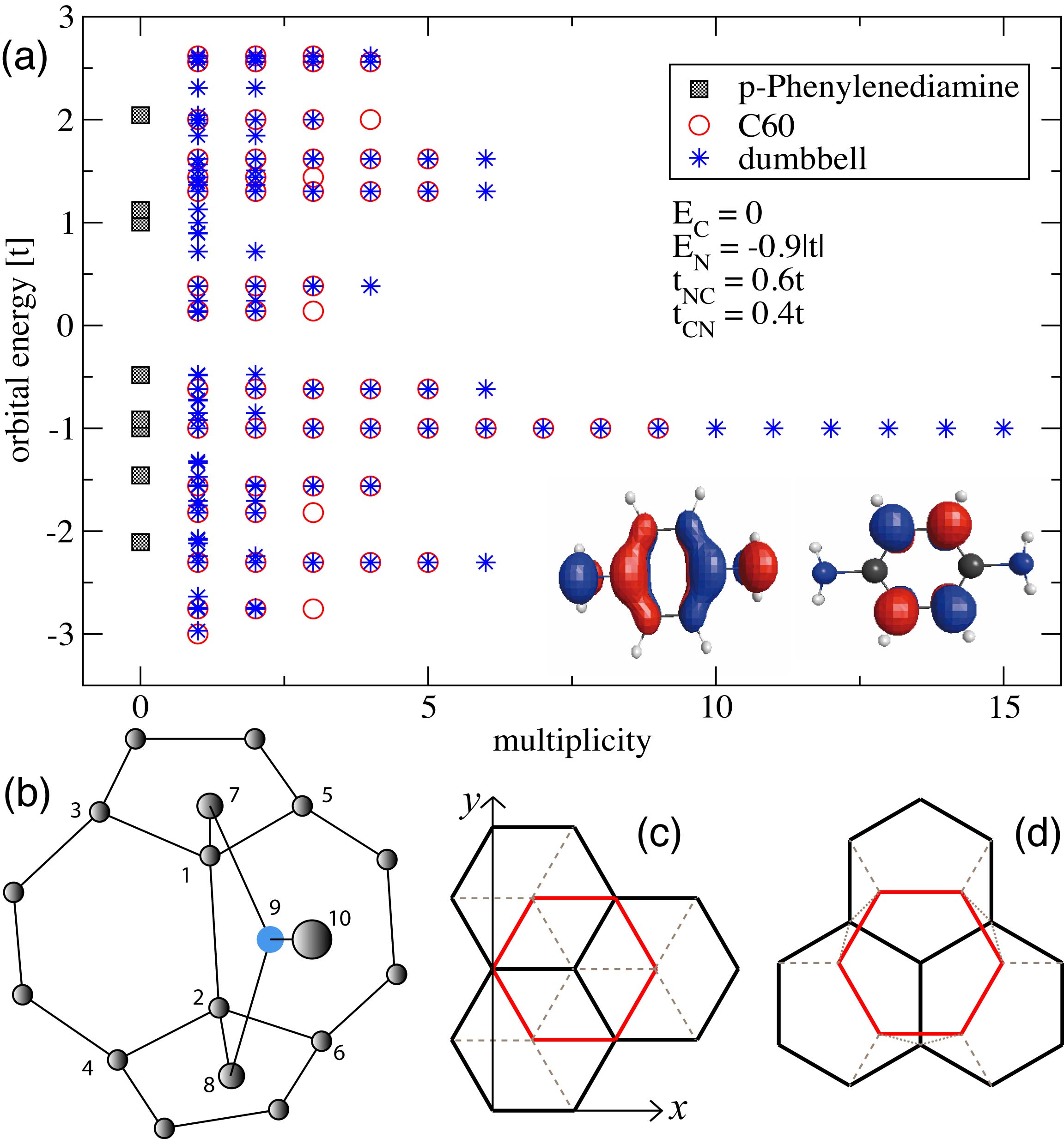}
  \caption{(a) Molecular eigenvalues of the phenylenediamine bridge
    without C60 end-groups (black squares), the C60 molecule (red
    rings), and the dumbbell molecule (blue stars) within the
    tight-binding model with the indicated parameters.  The dumbbell
    HOMO originates from the phenylenediamine bridge, while the LUMO
    originates from the end-groups. The insets show the HOMO (left)
    and LUMO (right) of the phenylenediamine bridge computed with the
    quantum chemistry package GAMESS.\cite{QC} The symmetries of these
    orbitals are also obtained within the tight-binding model. (b)
    Geometry of the molecular bridge. Carbon atoms number 1, 2, 7, and
    8 are in $sp^3$ orbital hybridization and do not participate in
    current transport, but we include next-nearest neighbor coupling
    $t_{CN}=0.4t$ between atoms 3, 4, 5, and 6 to the nitrogen atom
    (number 9). The hopping from nitrogen to the nearest neighbor in
    the benzene ring (atom number 10) is $t_{NC}=0.6t$. (c) Bernal
    stacking of a hexagon in C60 (red) and graphene (black);
    orientation O2 in Fig.~\ref{Fig-vdW}. (d) Stacking for a
    $30^{\circ}$ rotation of the C60; orientation O1 in
    Fig.~\ref{Fig-vdW}. The nearest neighbor hopping [vertical in (c)
    and dotted in (d)] and next-nearest neighbor hopping (dashed
    lines) between the two layers are $t_1=0.28t$ and $t_2=0.22t$,
    respectively.}
\label{Fig-EIG}
\end{figure}

For the molecule, the C60 end groups and the benzene ring of the
bridge are all carbon based, while each link group in addition
contains a nitrogen atom. For our purposes, it is enough to model the
molecule within the tight-binding (H\"uckel) theory on equal footing
with the leads, and leave details to be explored in future
calculations and experiments. The parameters of the molecule are kept
to a minimum by only varying hopping between C60 end groups and the
center phenylenediamine bridge, while letting all other carbon atoms
have the same parameters in the molecule as in the graphene leads. The
molecule parameters are the onsite energy on the nitrogen atom
relative to the onsite energy of the carbon atoms,\cite{Inga}
$E_N=E_C-0.9|t|$, ($t$ is the C-C hopping amplitude), the hopping from
C60 to the nitrogen group $0.4t$, and the hopping from the benzene
ring to the nitrogen group $0.6t$, see Fig.~\ref{Fig-EIG}(b). The
energy levels and orbitals of the C60 end groups are known within the
tight-binding model.\cite{Dresselhaus_book} For the bridge, we have
compared our model with orbitals obtained with the freely available
quantum chemistry package GAMESS,\cite{QC} see
Fig.~\ref{Fig-EIG}(a). We would like to emphasize that if the
parameters of this model are varied, unimportant details of the
results may change, but the general principles of how the transistor
operates will not change.

Finally, to determine the C$_{60}$-on-graphene adsorption geometry and
estimate both the magnitude of effective C$_{60}$-graphene hopping
constants and the C$_{60}$-graphene charge transfer, we have performed
a study using the van der Waals density functional (vdW-DF)
method.\cite{dion2004,thon2007} The results have been obtained in a
non-selfconsistent evaluation of the most recent version,
vdW-DF2,\cite{lee2010} a version which some of us have previously
found gives an accurate description of the binding in both a C$_{60}$
crystal and of graphene layers.\cite{ber2011} We have used a
plane-wave code and a standard semi-local density functional
approximation, to obtain underlying results for the electron-density
variation (as a function of adsorption geometries and distances). Our
choice of non-selfconsistent vdW-DF evaluations is motivated by a
recent analysis.\cite{thon2007}

\begin{figure}
  \includegraphics[width=\columnwidth]{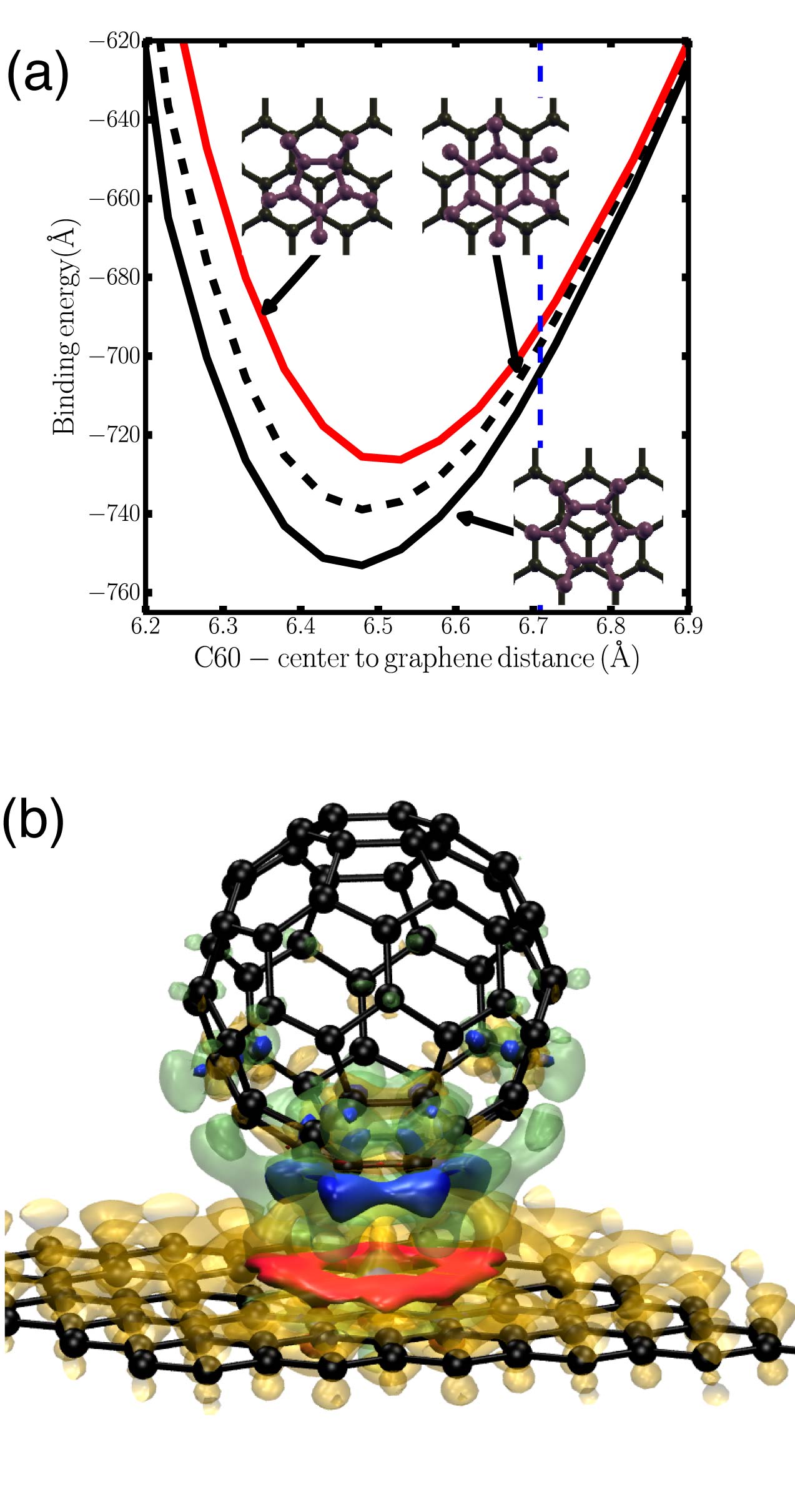}
  \caption{Results of a density functional theory calculation
    including the van der Waals interaction. (a) Binding energy of C60
    on graphene for three high-symmetry configurations. The solid
    black line (orientation O1) and dashed black line (orientation O2)
    correspond to dumbbell molecule anchoring for zigzag and armchair
    leads, respectively. In both cases, a hexagon of the C60 faces
    graphene. The configuration with a pentagon facing graphene (red
    line) is less favorable. (b) The charge density distribution at
    the C60-graphene contact for orientation O1. The blue (red) area
    is negative (positive) charge, implying a visible charge transfer
    to C60 from graphene.}
  \label{Fig-vdW}
\end{figure}

We present in Fig.~\ref{Fig-vdW} a summary of our vdW-DF study. Panel
(a) shows vdW-DF2 results for the variation in binding energy of
C$_{60}$ on graphene with center-of-mass separation for three typical
low-energy adsorption geometries that we have investigated in an
extended search.  We find that there is a systematic preference for
adsorption with the C$_{60}$ hexagon facing down and situated on top
of a graphene atom.  The panel also provides a comparison of this type
of adsorption geometries (identified by the inserts which shows
C$_{60}$ atoms in purple, graphene atoms in black) and we find that
the most favorable configuration corresponds to a 30$^{\circ}$
rotation of what would amont to a Bernal stacking of the hexagon on
the graphene. Important for the transport modeling, we predict that
the optimal adsorption separation is smaller than the value (vertical
dashed line) which would corresponds to the predicted layer separation
in graphite. We conclude that the effective C$_{60}$-to-graphene
hopping constants must be chosen larger than the choice which is made
in a tight-binding modeling of graphite; in the qualitative transport
modeling below, we simply set the enhancement at a factor of two.  The
two lowest energy configurations (solid and dashed lines in
Fig.~\ref{Fig-vdW}) corresponds, in the transport calculation below,
to anchoring of the dumbbell molecule for zigzag (orientation O1) and
armchair (orientation O2) leads, respectively [see
Fig.~\ref{Fig-EIG}(c)-(d)].

In Fig.~\ref{Fig-vdW}(b) we show the details of the C$_{60}$
adsorption and the complex charge transfer which we have calculated
for the optimal adsorption geometry. The blue colors shows regions of
electron accumulation whereas the red regions identify an electron
depletion (relative to a superposition of the graphene and C$_{60}$
electron densities). The panel shows that the binding is characterized
not only by van der Waals forces but also by a pronounced dipole (and
even multipole) formation. In addition, we find\cite{neat2006} a net
charge transfer from the graphene and to the C $_{60}$. We find that
the C$_{60}$-on-graphene binding is beyond simple physisorption as the
charge rearrangement causes a work function
shift.\cite{neat2006,kaas2008}

The first step in the transport characterization is to obtain the
retarded Green's function $G^R(E)$ of the system, which is a matrix in
the site indices. We utilize our own implementation of a recently
developed recursive algorithm\cite{waintal} within which the sites are
added one by one, which is ideal for our devices with complicated
geometries. The advanced Green's function is obtained by hermitian
conjugation $G^A(E)=\left[G^R(E)\right]^{\dagger}$. Observables are
related to the lesser Green's function $G^<$. In the absence of
electron correlations, the expression for the lesser Green's function
is reduced to the form
\begin{eqnarray}
G_{ij}^<(E) &=& \sum_{\ell} f_{\ell}(E) \nonumber\\
&&\times \sum_{c\tilde c} G_{ic}^R(E)
\left\{
\left[\Sigma_{\ell}^R(E)\right]^{\dagger}-\Sigma_{\ell}^R(E)
\right\}_{c\tilde c}
G_{\tilde cj}^A(E).\nonumber
%\label{Glesser}
\end{eqnarray}
It involves the distribution functions of the leads $f_{\ell}(E)$ and
self-energies $\Sigma_{\ell}^R(E)$ at the surfaces of the leads that
remain after eliminating the leads in favor of the system Green's
function. The leads are enumerated by the index $\ell$ (here $\ell=1$
and $2$ for source and drain) and surface sites of the leads are
labeled by $c$ and $\tilde c$. Local charge current flow in the device
(bond current between sites $i$ and $j$) is written as
\begin{equation}
I_{ij} = e\int_{-\infty}^{\infty}
\left[ t_{ij}G_{ji}^<(E)-t_{ji}G_{ij}^<(E) \right]dE.
\label{bondcurrent}
\end{equation}
The transmission function can also be written in terms of the retarded
Green's function and self-energies of the leads,
\begin{equation}
T(E) = \mbox{Tr}\left[ \Gamma_1(E) G^R_{12}(E) \Gamma_2(E) G^A_{21}(E) \right]
\end{equation}
where $\Gamma_{\ell}=i[\Sigma_{\ell}^R -
(\Sigma_{\ell}^R)^{\dagger}]$, $G^R_{12}$ is the propagator between
leads $1$ and $2$, and the trace is over the surface sites. Or we may
compute $T(E)$ by integrating the bond-currents
[Eq.~\ref{bondcurrent}] flowing through an interface of the device.

\section{Transistor effect}

\begin{figure}
 \includegraphics[width=\columnwidth]{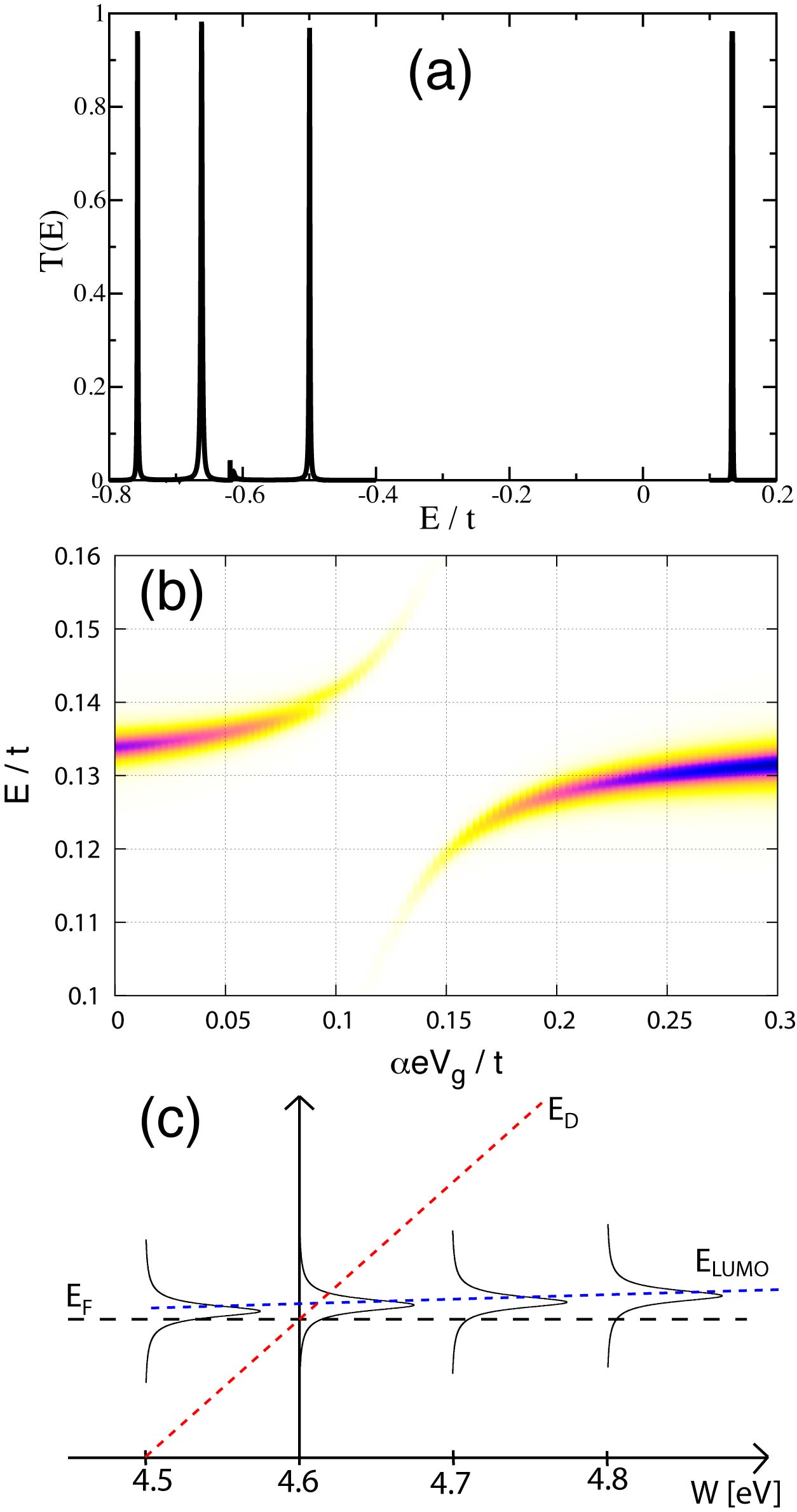}
 \caption{Electron transport through a single dumbbell molecule
   connected to armchair graphene leads with perfect edges. Anchoring
   of the dumbbell molecule in orientation O2.  (a) Transmission as
   function of energy when the Dirac point (here $E=0$) is below the
   LUMO, $E_{\mbox{\footnotesize LUMO}}\approx 0.133t$. (b)
   Transmission as function of energy and back-gate voltage for
   energies close to the LUMO. As the Dirac point is tuned by the
   back-gate through the molecular level, the transmission is quenched
   which leads to a transistor effect.  (c) A sketch of the movement
   of the Dirac point through the molecular level as the graphene lead
   work function is tuned by the back gate voltage. We estimate (see
   text) that the charge transfer to the molecule increases as the
   Dirac point $E_D$ in the graphene band structure is tuned to be
   below the Fermi level.}
\label{Fig-transistor}
\end{figure}

In Fig.~\ref{Fig-transistor}(a) we show an example of a transmission
function for one molecule in the center of the graphene nanogap in a
symmetric position, here for armchair leads. The transmission displays
typical resonance features near the molecular levels of the isolated
molecule. The levels are shifted and broadened by the coupling to the
leads. The amount of broadening depends on the exact coupling of the
molecule to the graphene, but also on the nature of the molecular
orbitals. For this particular molecule, the LUMO is mainly centered on
the C60 anchoring groups that act as effective extensions of the
leads, while the bridge acts as the weak link in the
system. Functionality can be added to the device by choosing a
different bridge by exchanging the benzene ring during molecular
synthesis.\cite{BDC_JOC} But here we shall continue working with the
benzene bridge and focus on the transistor effect.

In Fig.~\ref{Fig-transistor}(b) we present a contour plot of the
transmission function for energies (vertical axis) near the LUMO as we
rigidly shift the band structure of the graphene leads relative the
molecular level by a back gate voltage (horizontal axis with a
transfer function $\alpha$ between the gate voltage and the shift of
the Dirac point). The gate effect we have in mind can be visualized
starting from Fig.~\ref{Fig-ELD} as moving the graphene bands
vertically keeping molecular levels and the Fermi level fixed. When
the Dirac point is far from the molecular level, the transmission
resonance corresponds to the transistor in the on-state. As the Dirac
point in the bandstructure approaches the molecular level, either from
below or above in energy, the transmission resonance is shifted due to
hybridization with zigzag nanogap edge states. The possibility of such
hybridization was also noted recently in a DFT
calculation.\cite{mot2011}. When the Dirac point passes through the
level, the transmission is suppressed and the transistor is in the
off-state, see Fig.~\ref{Fig-transistor}(b) at $\alpha
eV_g=E_{\mbox{\footnotesize LUMO}}\approx 0.133t$. The on and off
states will be well separated when the Dirac point can be shifted by
$\delta E_D>\Gamma$, larger than the molecular level broadening
$\Gamma$.

The transistor on-off ratio will be large when the Fermi level $E_F$
is aligned close to a molecular level. This is the case here, with
$E_F$ in the broadened LUMO. We can estimate the molecular level
alignment\cite{Datta_Review} with respect to the Fermi level by
estimating the charge transfer between graphene and C60. We do that by
comparing the work function of graphene with metals for which the
charge transfer to C60 has been measured. It has been shown by
scanning Kelvin probe microscopy\cite{Yu_NL2009} that the application
of a back gate voltage results in a change of the graphene work
function between 4.5~eV (electron doped) and 4.8~eV (hole
doped). Scanning tunneling experiments and DFT calculations of C60 on
gold and silver surfaces show\cite{Lu_PRB2004} that the charge
transfer to C60 from gold is vanishingly small, while it is of order
$0.2e$ from silver. The work function of silver is 4.6~eV, while that
of gold is 5.3~eV. This picture is corroborated by our vdW-DF study
which identified a net charge transfer and dipole formation. In
summary, we draw the conclusion that there is considerable
charge-transfer effects already for hole-doped leads ($W=4.8$~eV),
which results in the Fermi level energy in our system aligned inside
the broadened LUMO. As the gate voltage is changed, the charge
transfer to the molecule will increase as we go through the Dirac
(neutrality) point of graphene to the electron doped side (eventually
reaching $W=4.5$~eV), which results in a Fermi level deeper inside the
LUMO.  Based on these estimates, where the Dirac point can be shifted
by at least $0.1t$ through the LUMO, see Fig.~\ref{Fig-transistor}(d),
the on-off ratio will be large but the precise value will ultimately
have to be determined by experiment. The back gate is straightforward
to use compared with the traditional direct gating of the molecule
itself. In fact, the transistor effect is most effective when the gate
is decoupled from the molecule and only affects the graphene leads.

\begin{figure}
 \includegraphics[width=\columnwidth]{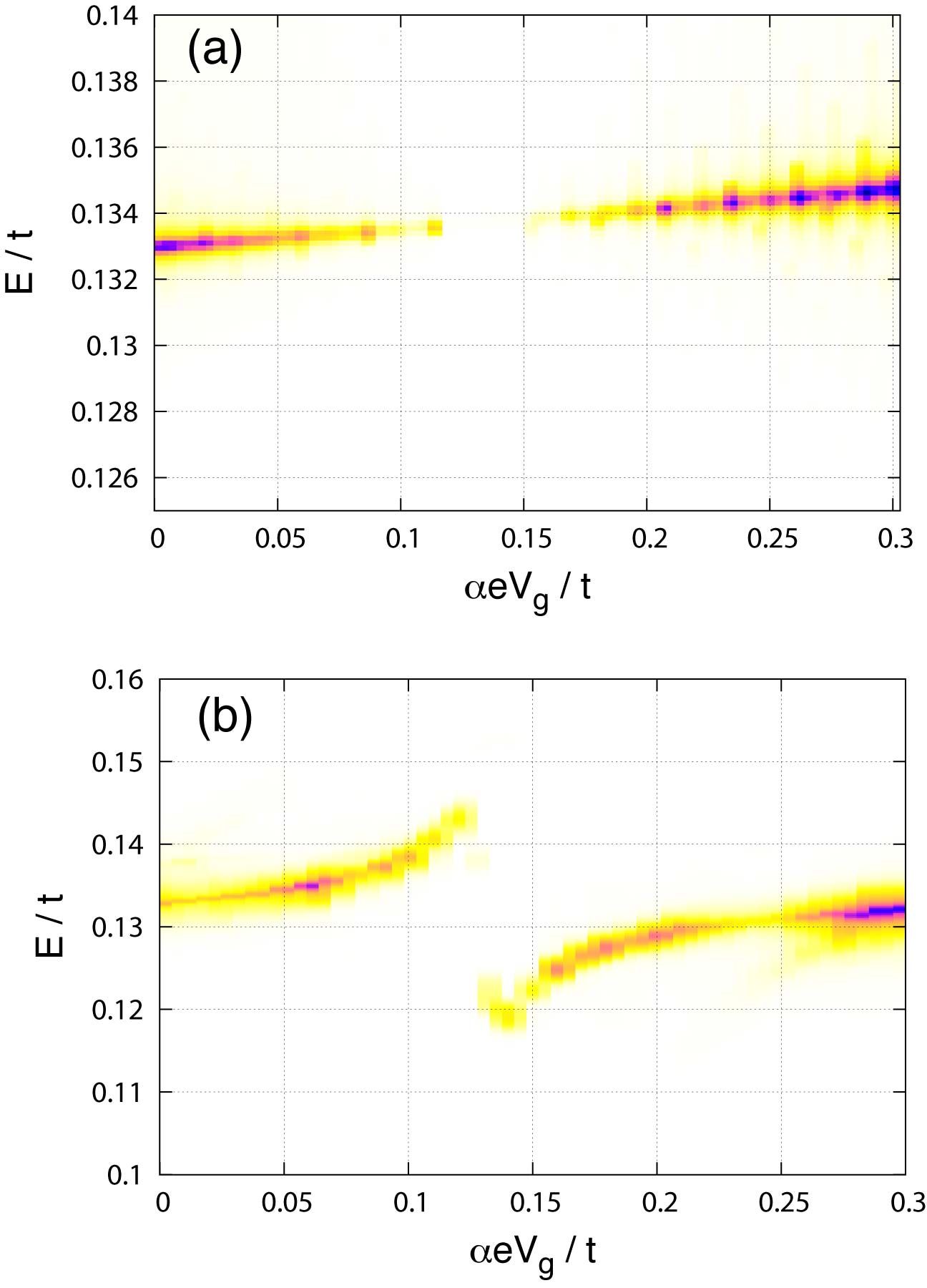}
 \caption{Electron transport through a single dumbbell molecule
   connected to graphene leads. (a) Zigzag leads with perfect edges;
   anchoring of the dumbbell molecule in orientation O1. (b) Armchair
   leads with random rough edges; compare Fig.~\ref{Fig-transistor}(b)
   for perfect edges.}
\label{Fig-transistor2}
\end{figure}

In Fig.~\ref{Fig-transistor2}(a) we show the gate effect for the case
of wide zigzag graphene leads, with the anchoring of the C60 end
groups in orientation O1. For this orientation, there is no nanogap
edge states, since the nanogap has armchair orientation. The strong
hybridization of the molecular level with lead states is therefor
absent and we predict a simple weak shift of the molecular level with
gate voltage. As the Dirac point is tuned through the molecular level,
the transmission is quenched, and we have a transistor effect
analogous to the case with armchair leads discussed above.

We note that graphene itself (without nanogap and molecules as weak
links) works as a transistor via the back gate. However, the graphene
transistor can not be set in the off-state as the minimal conductivity
is of order $e^2/h$, in contrast to the molecular transistor with
graphene leads that we study here. A nanoribbon has an energy gap
related to its width, and would work as a transistor with an
off-state.  However, the required ribbon width is small (a few nm) and
it is very difficult to control the nanopatterning with the required
atomic resolution. In contrast, as we show below, the molecular
graphene nano-gap transistor is more robust against edge disorder in
the graphene leads.

\section{Charge flow patterns}
 
Since graphene is 100~\% surface, it is an ideal material to study by
scanning techniques.\cite{con2010} Previously, atoms and molecules on
metal surfaces\cite{Lu_PRB2004} or large-area graphene\cite{bra2011}
have been studied by scanning tunneling microscopy (STM) and
spectroscopy (STS) and valuable details about e.g. orbitals and charge
transfer have been obtained. Scanning techniques have been utilized to
reveal local information about Coulomb interactions in graphene
nanostructures.\cite{schnez10} Quantum transport through quantum point
contacts in 2DEGs have been mapped out\cite{Topinka_review} by
scanning gate spectrocsopy (SGS) and revealed so-called branched flow
originating from a background random potential due to doping
impurities in the layers forming the 2DEG. SGS has also been used to
reveal coherent electron transport in large-area graphene
flakes.\cite{ber2010a,ber2010b} Clearly, transport in a molecular
device with metal electrodes is hidden by the bulky nature of the
metallic contacts. Graphene on the other hand, being two-dimensional,
would be a uniquely suitable electrode enabling information about
quantum transport in a molecular device to be revealed by scanning
techniques.

\begin{figure}
  \includegraphics[width=\columnwidth]{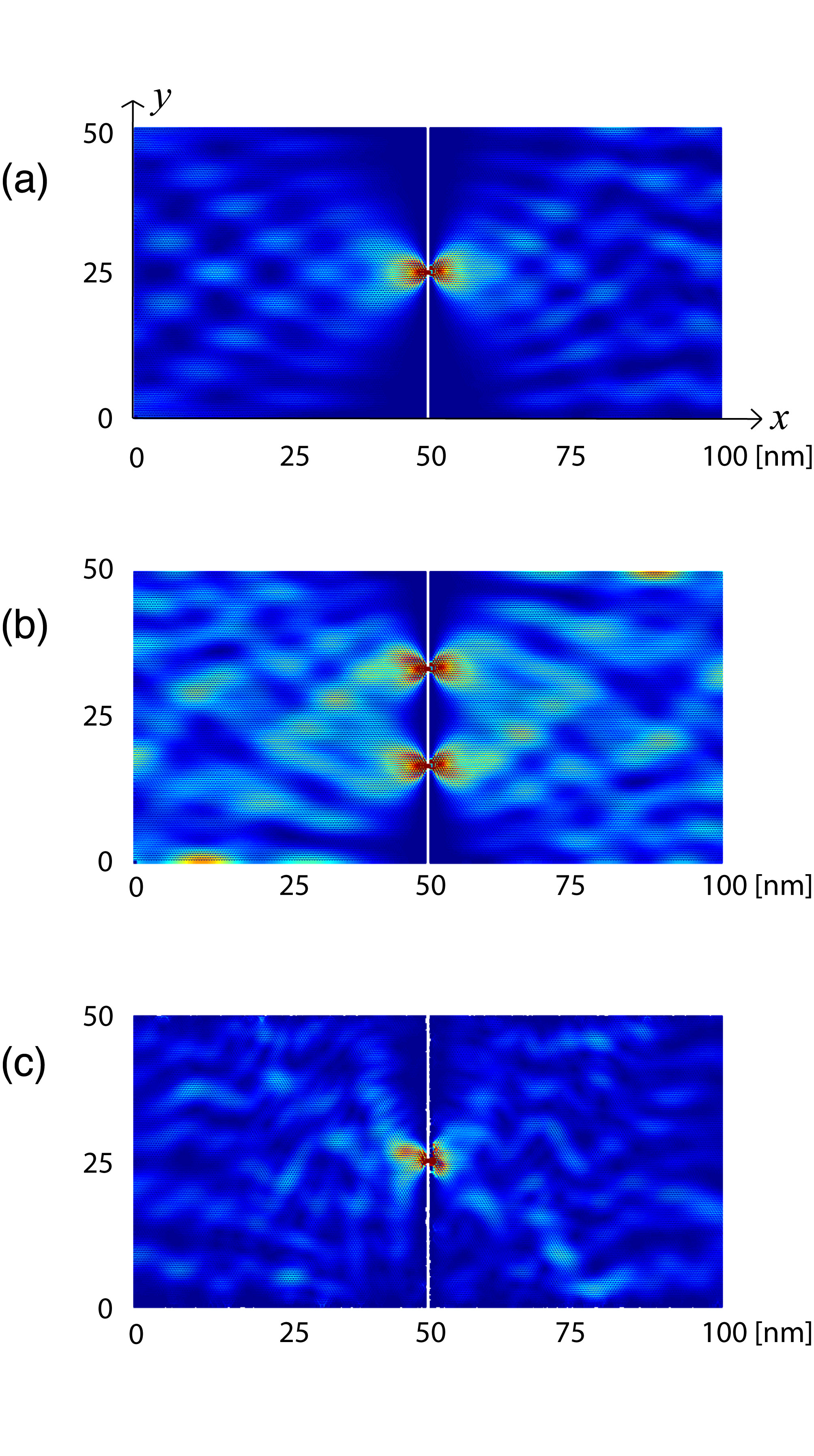}
  \caption{(a) Spectral current flow pattern through a transistor
    with one molecule in the nanogap. Note that the molecule is
    much smaller than the lobe structured pattern in the charge flow
    in and out of the molecule visible in the leads. (b) Spectral
    current flow pattern for the case of two molecules in the nanogap.
    (c) Spectral current flow patterns disturbed by random edges of
    the leads. In all cases, we assume zero temperature and study the
    linear low-voltage regime.}
\label{Fig-flow}
\end{figure}

In Fig.~\ref{Fig-flow}(a) we present the spectral current flow pattern
[the integrand in Eq.~\ref{bondcurrent}] through the device for an
energy corresponding to the top of the LUMO peak in the transmission
shown in Fig.~\ref{Fig-transistor}(a). The position of the molecule is
clearly visible in the current flow pattern. The molecule forms the
weak link where all current is channeled through. The current is
flowing in and out of the molecule in a characteristic lobe pattern,
that is due to the specific anchoring of the C60 on graphene. Deep
inside the electrodes, the current is carried throughout the width of
the ribbon.

In Fig.~\ref{Fig-flow}(a) the molecule is in a symmetric position in
the nanogap. If the molecule is in an asymmetric position, the charge
flow pattern is simply displaced vertically and changes in an
intuitive way (not shown) and the transmission function remains
unchanged unless the molecule is very close to the upper or lower
edges of the nanogap, within a few rings in the graphene
leads. Similarly, if two molecules are bridging the nanogap, to a good
approximation two flow patterns are simply superimposed, as we show in
Fig.~\ref{Fig-flow}(b). We may expect quantum intereference between
the two pathways (two molecules).  The effect on the transmission
function is weak in this example, however, unless the two molecules
are very close to each other, with the anchoring groups only separated
by a few rings in the graphene leads.

In a real device, perfect armchair or zigzag edges are hard to
fabricate. In Fig.~\ref{Fig-flow}(c) we show an example of the effect
of edge disorder, consisting of randomly removed carbon atoms within
one ring from the original perfect edges. The interference patterns in
the leads are now affected, but the molecular weak link remains
clearly visible in the patterns of current flowing in to and out of
the molecule. Also the transmission function $T(E)$ is in its main
features unaffected by defects in the leads, as we show in
Fig.~\ref{Fig-transistor2}(b) [compare
Fig.~\ref{Fig-transistor}(b)]. Ideal graphene leads are not crucial
for the transistor to function, since the weak link is the molecule.

\section{Summary and Conclusions}

In conclusion, we have studied a single molecule device with graphene
leads working in the quantum coherent transport regime. We predict a
transistor effect that is pronounced when the gate coupling to the
leads is much stronger than to the molecule. This opens new avenues
for research of gate tunable quantum coherent molecular electronics
with single molecules.

\section*{Acknowledgement}

We would like to thank V. Geskin and G. Wendin for valuable
discussions at various stages of this work. This work has been
supported by the European Union through the projects SINGLE and
ConceptGraphene, as well as SSF, the Swedish foundation for strategic
research.

\end{document}